\documentclass[aps,prb,twocolumn,superscriptaddress]{revtex4-1}
\usepackage{amsmath}
\usepackage[pdftex]{graphicx}
\usepackage[thinspace]{SIunits}

\providecommand \BibitemShut  [1]{\csname bibitem#1\endcsname}%

\begin{document}


\title{Excited State Quantum Couplings and Optical Switching of an Artificial Molecule}

\author{K. M\"uller}
 \affiliation{Walter Schottky Institut, Technische Universit\"at M\"unchen, Am Coulombwall 4, 85748 Garching, Germany \\}
\author{G. Reithmaier}
 \affiliation{Walter Schottky Institut, Technische Universit\"at M\"unchen, Am Coulombwall 4, 85748 Garching, Germany \\}
\author{E. C. Clark}
 \affiliation{Walter Schottky Institut, Technische Universit\"at M\"unchen, Am Coulombwall 4, 85748 Garching, Germany \\}
\author{V. Jovanov}
 \affiliation{Walter Schottky Institut, Technische Universit\"at M\"unchen, Am Coulombwall 4, 85748 Garching, Germany \\}
\author{M. Bichler}
 \affiliation{Walter Schottky Institut, Technische Universit\"at M\"unchen, Am Coulombwall 4, 85748 Garching, Germany \\}
\author{H. J. Krenner}
 \affiliation{Lehrstuhl Experimentalphysik 1 Universit\"at Augsburg 86159 Augsburg \\}
 \author{M. Betz}
 \affiliation{Experimentelle Physik 2, TU Dortmund, 44221 Dortmund, Germany \\}
\author{G. Abstreiter}
 \affiliation{Walter Schottky Institut, Technische Universit\"at M\"unchen, Am Coulombwall 4, 85748 Garching, Germany \\}
\author{J.J. Finley}
 \email{finley@wsi.tum.de}
 \affiliation{Walter Schottky Institut, Technische Universit\"at M\"unchen, Am Coulombwall 4, 85748 Garching, Germany \\}

\date{\today}

\begin{abstract}
We optically probe the spectrum of ground and excited state transitions of an individual, electrically tunable self-assembled quantum dot molecule. Photocurrent absorption measurements show that the spatially direct neutral exciton transitions in the upper and lower dots are energetically separated by only $\sim2$ meV. Excited state transitions $\sim8-16$ meV to higher energy exhibit pronounced anticrossings as the electric field is tuned due to the formation of hybridized electron states.  We show that the observed excited state transitions occur between these hybridized electronic states and different hole states in the upper dot. By simultaneously pumping two different excited states with two laser fields we demonstrate a strong ($88\%$ on-off contrast) laser induced switching of the optical response. The results represent an electrically tunable, discrete coupled quantum system with a \textit{conditional} optical response.
\end{abstract}


\maketitle



Quantum dot (QD) nanostructures formed by strain driven self-assembly are ideal for solid state quantum optics experiments due to their discrete optical spectrum, strong interaction with light and robust quantum coherence for both interband polarization\cite{Borri01,Borri03} and spin\cite{Greilich09}. The ease with which such nanostructures can be embedded into electrically active devices allows for tuning of the transition frequency and control of charge occupancy \cite{Warburton00}. Self-assembly provides a natural way to realize few dot systems via vertical stacking to produce more sophisticated nanostructures with coherent inter-dot coupling due to carrier tunneling\cite{Stinaff2006, PhysRevLett.97.076403, Doty2008, Kim2010,PhysRevLett.94.057402, Bracker2006, Scheibner2007biex, Scheibner2008}.  When combined with the potential to coherently manipulate excitons over ultrafast timescales using precisely timed laser and electrical control pulses \cite{Zrenner02,Zecherle2010,Vasconcellos10} such systems raise exciting prospects for the operation of small scale few qubit systems in a solid-state device. Very recently, conditional quantum dynamics for a single resonantly driven QD-molecule (QDM)\cite{Robledo2008} and spin dependent quantum jumps have been observed\cite{Kim2010, Atatuere2010}.  

In this paper we employ photocurrent (PC) absorption, photoluminescence (PL) emission and PL-excitation (PLE) spectroscopy to trace the spectrum of ground and excited state transitions of an individual self-assembled QD-molecule as their character is electrically tuned from spatially direct to indirect. PC absorption allows us to identify the spatially direct neutral exciton transitions in both the upper ($X_{ud}$) and lower ($X_{ld}$) dots in the molecule.  A number of excited state transitions  are identified in PLE $\sim8-16$ meV above $X_{ud}$.  These excited states exhibit pronounced anticrossings (energy splitting $\Delta E \sim 3.2-3.5$ meV) as the electric field $F$ is tuned. Excited state transitions are identified from voltage dependent PLE measurements to correspond to transitions between these hybridized electronic states and different \textit{hole} orbitals in the upper dot. By performing a multi-color experiment where the QDM is simultaneously excited with different frequency lasers, we demonstrate how the resonant excitation of indirect excitons or excitons in the lower quantum dot can be used to suppress the resonant excitation of the upper quantum dot, due to interdot Coulomb interactions. An on-off gating contrast up to 88$\%$ is observed, demonstrating a \textit{conditional} optical response of an artificial molecule.

\begin{figure}
\includegraphics[width=1\columnwidth]{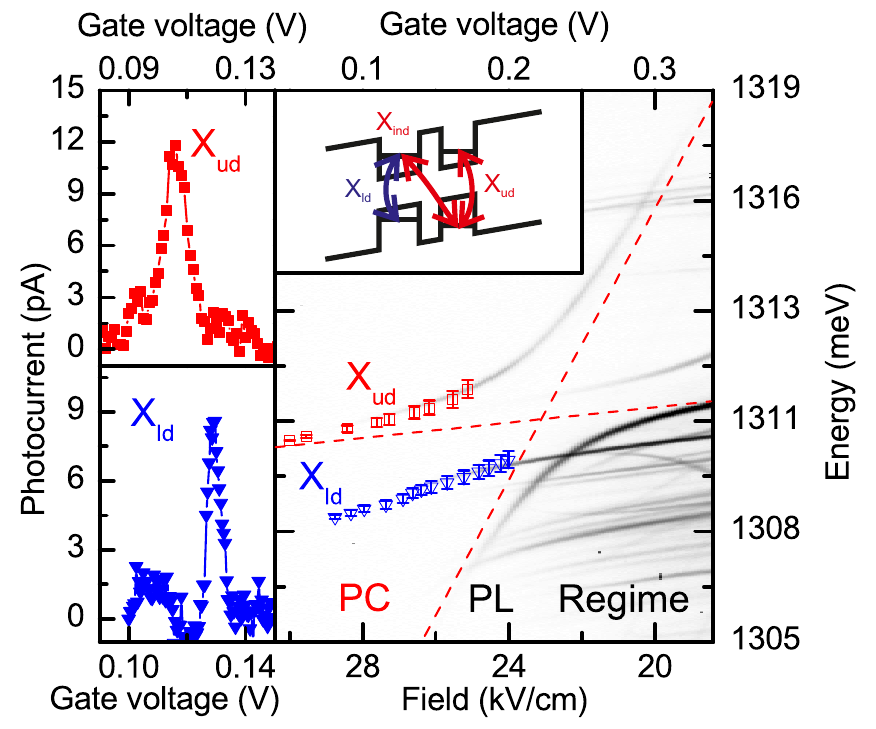}
\caption{\label{fig:Figure_1}
(Color online) 
Combined result of an electric field dependent PL and PC measurement: The PL of the QDM as a function of the applied electric field is shown as a contour plot from 0 to 130cps from white to black on a logarithmic scale. The dashed red lines indicate the uncoupled neutral exciton in the upper dot and the indirect exciton as depicted schematically in the inset. The PC resonances are shown as red squares and blue triangles for the neutral exciton transitions in the upper and lower dot, respectively.}
\end{figure}

The sample consists of vertically stacked pairs of QDs separated by a 10nm thick GaAs spacer and embedded within the intrinsic region of a GaAs Schottky photodiode \cite{PhysRevLett.94.057402}. Typical PL and PC measurements recorded at T=4.2K are presented in Fig.1. For the PL measurement the sample was excited in the wetting layer at 1.49 eV. Typical electric field dependent PL from  18 to 32 kV/cm are presented in a greyscale contour plot representation in Fig.1. The measurements show an anticrossing of two transitions arising from spatially direct and indirect excitons in the QDM where the hole is located in the upper dot \cite{PhysRevLett.94.057402}. These optical transitions are depicted schematically in a single particle picture in the inset of Fig.1: the direct exciton in the upper dot $X_{ud}$ and the indirect exciton $X_{ind}$ with the hole in the upper dot and the electron in the lower dot. The indirect exciton exhibits a strong Stark shift due to the large static dipole. As the energies of the two states are tuned to resonance, electron mediated  tunnel coupling occurs that results in the formation of molecular bonding (lower energy) and antibondig (higher energy) orbitals \cite{ab-b} and the observed anticrossing \cite{PhysRevLett.97.076403}. As $F$ increases beyond $\sim 25$ kV/cm, the intensity of the luminescence reduces as charge carriers escape from the QDM via tunneling and PC measurements can be performed with resonant optical excitation. Two prominent resonances are observed in PC, examples of which are presented in the left panel of Fig.1. The energy of these peaks are plotted in the main panel of Fig.1 for F=25-31 kV/cm. The transitions observed in PC arise from charge neutral excitons and show that the QD molecule exhibits \textit{two} neutral exciton absorption resonances.  As $F$ decreases these two resonances clearly evolve into the two clear peaks observed in PL, labeled $X_{ld}$ and $X_{ud}$ on Fig.1. The state marked as $X_{ld}$ is attributed to an exciton in the lower quantum dot of the molecule (depicted in blue in the inset of Fig.1) while $X_{ud}$ is the direct exciton in the upper dot, assignments that are confirmed by the results presented below.

\begin{figure}
\includegraphics[width=1\columnwidth]{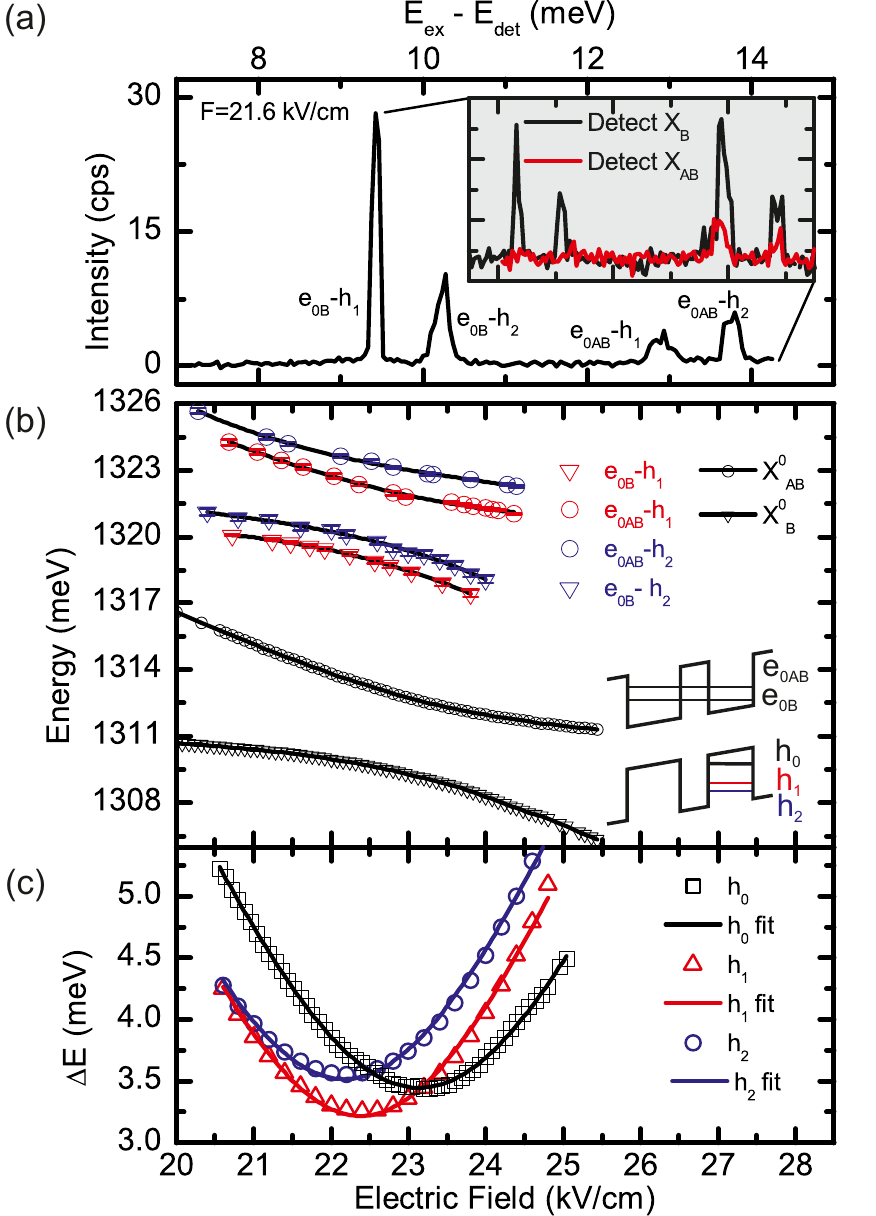}
\vspace{-15pt}
\caption{\label{fig:Figure_2}
(Color online) 
(a) Typical PLE scan detecting on $X_{B}$. The inset compares PLE recorded at F= 23.2kV/cm detecting either $X_B$ (black) or $X_{AB}$ (red), respectively. 
(b) Spectrum of QDM transitions determined by PL (black symbols) and PLE (blue, red symbols) States with bonding (antibonding) character are plotted as triangles (circles). (c) Energy separation ($\Delta E$) between corresponding anticrossing states, fits using equation \ref{the_equation}.}
\end{figure}

We conducted detailed PLE measurements to track the evolution of the excited state spectrum of the QDM as a function of $F$.
A typical PLE scan detecting on $X_{ud}$ is presented in Fig.2(a) with $F$ fixed close to the anticrossing (21.6 kV/cm). Several discrete electronic resonances are observed in this region, the first four of which are labeled $e_{0B}-h_1$, $e_{0B}-h_2$, $e_{0AB}-h_1$ and $e_{0AB}-h_2$, in Fig.2(a). This assignment anticipates the nature of these excited states corresponding to the electron being in the bonding (B) or antibonding (AB) molecular orbital while the hole occupies the first ($h_1$) or second ($h_2$) excited orbital state in the upper QD. These assignments are now justified by examining the electric field dependence of the excited state resonances.  
Fig.2(b) shows the energy of the molecular ground states determined via PL and the first four excited states as a function of $F$ in the range 20-25 kV/cm measured using PLE. The first four excited states consist of two different pairs of lines colour coded by the red and blue symbols in Fig.2(b), each of which anticross at an electric field close to 22 kV/cm. To analyze the observed excited state anticrossings in more detail and compare to the anticrossing of $X_{B}$ and $X_{AB}$ observed in PL, we plot the energy separation ($\Delta E$) between bonding and antibonding state of the ground states and the two excited states anticrossings in Fig.2(c). Thereby, the energy difference is plotted using the same colour coding as the corresponding anticrossing in Fig.2(b).

\begin{table}
\begin{tabular}{|c|c|c|c|}
\hline
Hole state & $2V_{ee}$ (meV) & $F_0$ (kV/cm) & d (nm) \\ \hline
$h_0$ & $3.4 \pm 0.1$ & $23.1 \pm 0.1$ & $15.3 \pm 0.1$ \\ \hline
$h_1$ & $3.2 \pm 0.1$ & $22.4 \pm 0.1$ & $15.8 \pm 0.2$ \\ \hline
$h_2$ & $3.5 \pm 0.1$ & $22.2 \pm 0.1$ & $15.9 \pm 0.2$ \\ \hline
\end{tabular}
\caption{Results of the fits of $\Delta E$ from Fig.2(c) with eq.1}
\label{table1}
\end{table}

For  all three anticrossings $\Delta E$ shows a similar hyperbolic behaviour that can be fitted with
\begin{equation}
\Delta E = \sqrt{(2V_{ee})^2 + (ed(F-F_0))^2}
\label{the_equation}
\end{equation}
where $V_{ee}$ denotes the interdot tunnel coupling strength, $F_0$ the field at which the states anticross and $ed$ is the equivalent static dipole moment of the indirect exciton. ($d$ is the distance between the centers of the electron and hole envelope functions.) Fits to the three anticrossings are presented as lines in Fig.2(c) and the extracted values of $F_0$ and $d$ are summarized in table \ref{table1}. For each anticrossing, the extracted values of $d$ vary only slightly and are fully consistent with the expected electron-hole separation for the indirect exciton, since the dots have a typical height of 5 nm and a nominal separation of 10 nm for this sample. Both $V_{ee}$ and $F_0$ remain unchanged for the transitions involving $h_0$, $h_1$ and $h_2$, providing evidence that they arise from transitions between the same electron mediated anticrossing and different hole levels. This expectation is confirmed by PLE measurements performed close to $F_0$, detecting on either $X_{AB}$ or $X_B$ respectively.  Typical results are presented in Fig.2(a)(inset).  When detecting on $X_{AB}$ (red curve),  transitions are only observed for $e_{0AB}-h_1$ and $e_{0AB}-h_2$, whilst $e_{0B}-h_1$ and $e_{0B}-h_2$ are absent.  This arises since the electron populates the lower energy bonding level and thermal activation into the higher energy bonding level is unlikely since $2V_{ee}>>k_BT$ \cite{Nakaoka2006}.  In contrast, upon exciting states with bonding electron character all four resonances are observed due to phonon mediated thermalization from antibonding to bonding electron states.\cite{Nakaoka2006}  The small differences between $F_0$ arise from the Coulomb interactions between the various hole orbital states. Compared to the ground state with $F_0 = 23.1\pm0.1$ kV/cm the critical field of the excited state transitions are shifted by $-0.7\pm0.2$ kV/cm and $-0.9\pm0.2$ kV/cm. This can be converted to an energy difference of $1.1\pm0.3$ meV and $1.4\pm0.4$ meV using $d=15.8 \pm 0.2$ nm and $d=15.9 \pm 0.2$ nm, respectively. These values correspond to less than $10 \%$ of the total attractive e-h Coulomb interaction that has been estimated to be $\sim 22$ meV for similar samples.\cite{PhysRevLett.94.057402} Thus, the Coulomb shifts between the different valence band states represent a small pertubation and we conclude that these observations provide strong evidence that the four excited states shown in Fig.2(b) take place between an electron in either the lowest energy bonding or antibonding levels and different hole states in the upper dot, as depicted in the inset of Fig.2(b). The energy splitting between the lowest energy hole state and the first two excited states is 9.5 and 10.5 meV.

\begin{figure}
\includegraphics[width=1\columnwidth]{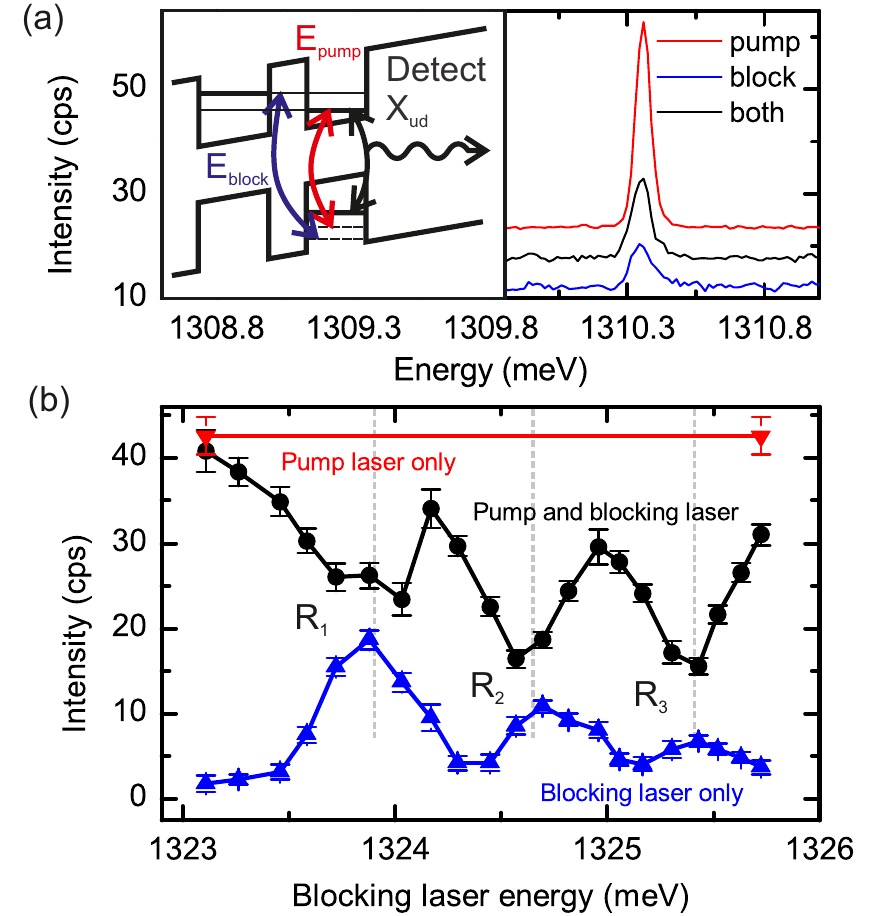}
\caption{\label{fig:Figure_3}
(Color online) 
(a) PL intensity of $X_{ud}$ for the excitation of a resonance with direct character (pump, red), indirect character(block, blue) and both lasers (black). PL from the excitation of the direct resonance is quenched due to the presence of the block laser.
(b) Intensity of $X_{ud}$ as a function of the block laser energy. Whenever the block laser hits an excited state with indirect character the PL from the upper dot is decreased.}
\end{figure}

Away from resonance for $F < F_0$ the bonding ground state as well as the first two excited states $e_{0B}-h_1$ and $e_{0B}-h_2$ have predominant \textit{direct} character, while the third and fourth excited states $e_{0AB}-h_1$ and $e_{0AB}-h_2$ have \textit{indirect} character. Therefore, exciting $e_{0B}-h_1$ and $e_{0B}-h_2$ is expected to populate $X_{ud}$ whilst exciting $e_{0AB}-h_1$ and $e_{0AB}-h_2$ generates an indirect exciton $X_{indir}$. We devised an experimental scheme using the direct and indirect character of the excited states to test whether the QDM exhibits a conditional optical response. The inset of Fig.3(a) illustrates schematically the principle of the measurement. $F$ is chosen such that $e_{0B}-h_1$ and $e_{0B}-h_2$ have predominant \textit{direct} excitonic character and $e_{0AB}-h_1$ and $e_{0AB}-h_2$ predominantly \textit{indirect} character. The system is resonantly excited by either one laser or two lasers simultaneously. In the two laser experiment, the first laser termed \emph{pump} is resonant with $e_{0B}-h_1$, as indicated by the red arrow  on the inset of Fig.3(a). As discussed above, an exciton created by laser absorption will primarily relax to the $X_{ud}$ ground state before PL is measured via phonon mediated processes \cite{PhysRevB.54.11532}. A second laser, termed \emph{block}, is tuned into resonance with excited states that have a predominantly \textit{indirect} character in order to generate excitons with indirect character $X_{ind}$. If the absorption of the blocking laser is more efficient than that of the pump then the QDM will be driven into an indirect exciton state and the absorption of the pump laser is suppressed since absorption shifts to a biexcitonic state of the system. Typical results of such a measurements are presented in Fig.3(a) that shows the PL spectrum recorded from $X_{ud}$ for $F=21.05$ kV/cm when the pump laser only (0.5 kW/cm$^2$) is applied (red curve). This is compared with the situation when the system is excited only by the blocking laser (5 kW/cm$^2$ - blue curve) and when both lasers are applied simultaneously (black curve).  

\begin{figure}[t]
\includegraphics[width=1\columnwidth]{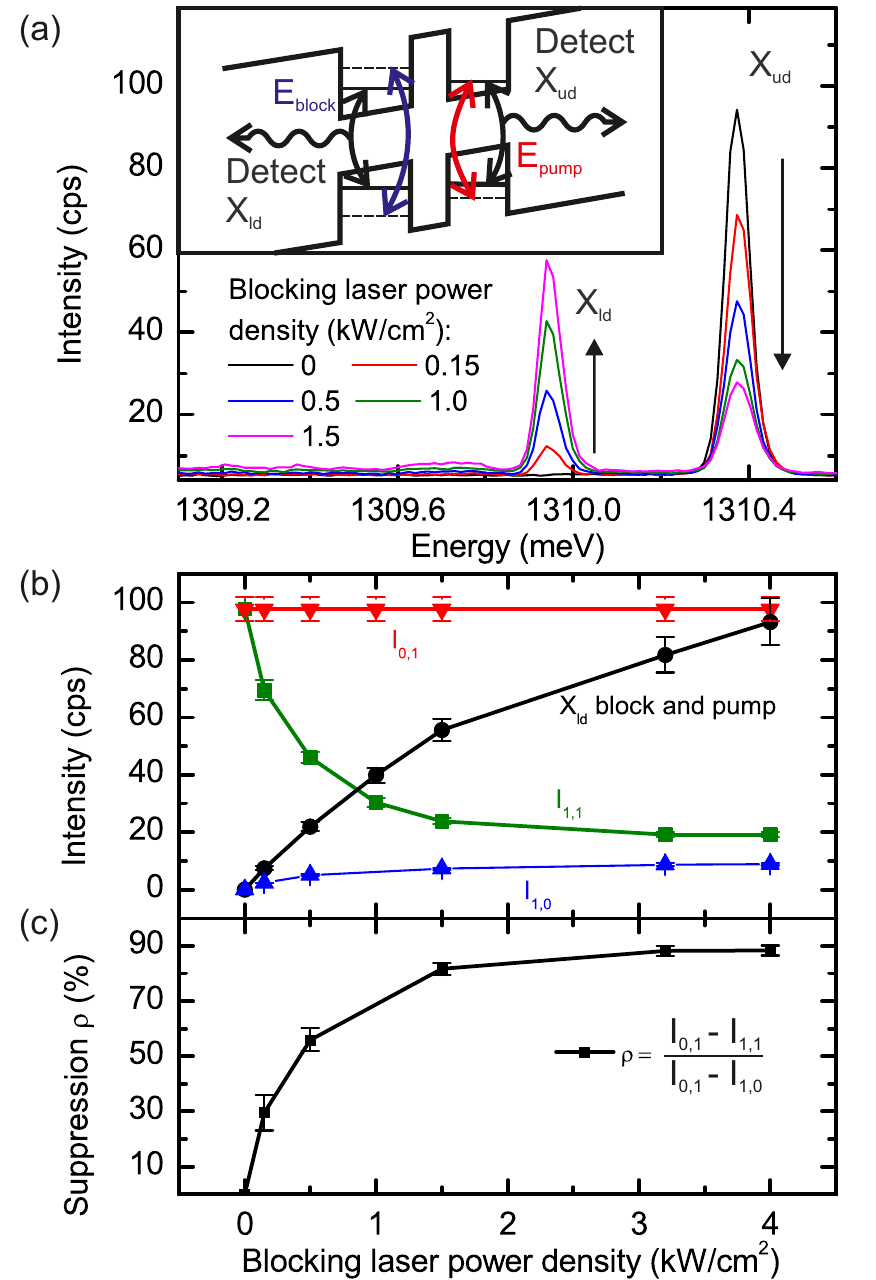}
\caption{\label{fig:Figure_4}
(Color online) 
PL of the upper and lower quantum dot as a function of the blocking laser power.
Resonant excitation of an exciton in the upper dot is blocked by the presence of an exciton in the lower dot due to the Coulomb interaction. (b) Intensity of the PL of the upper and lower dot and (c) suppression ratio $\rho$ as a function of the blocking laser power.}
\end{figure}

The intensity of $X_{ud}$ clearly reduces strongly when both lasers are applied simultaneously. We scanned the energy of the blocking laser over the spectrum of excited states from 1322 to 1326 meV. The result of this experiment is shown in Fig.3(b) comparing the intensity of $X_{ud}$ as a function of the blocking laser energy for the pump laser only (red curve), the blocking laser only (blue curve) and with both blocking and pump lasers applied simultaneously (black curve). Three resonances, labeled $R_1$, $R_2$ and $R_3$ in Fig.3(b) can be clearly seen. At these resonances PL is observed from $X_{ud}$ following excitation with the blocking laser only. The PL signal for excitation with both lasers shows a series of dips for an excitation with the pump laser only. $R_1$ and $R_2$ coincide precisely with the PLE resonances $e_{0AB}-h_1$ and $e_{0AB}-h_2$, presented in Fig.2(b) demonstrating that the blocking laser can be used to suppress the absorption of the pump beam.  

After establishing the spectrum of interband transitions of the molecule and their electric field dependence we performed an experiment whereby an excitation in one of the two dots forming the molecule was used to block absorption in the other. The scheme for this experiment is illustrated schematically in the inset of Fig.4(a). An exciton with predominantly direct character can be excited in the upper dot via its first excited hole state (red arrow). If, in addition to the pump laser the blocking laser is tuned to a direct exciton transition in the lower dot (blue arrow) the energy of the upper dot absorption shifts to that of the spatially separated biexciton.  Thus, the absorption of the pump laser is switched off and, thus, PL from $X_{ud}$ can be optically gated on and off. The result of such a measurement is presented in Fig.4(a) that shows the PL intensity of $X_{ud}$ and $X_{ld}$ for $F=20.8$ kV/cm, a pump laser power density of 1 kW/cm$^{2}$ and different blocking laser power densities from 0 to 1.5 kW/cm$^{2}$.

In the absence of the blocking laser, only PL from $X_{ud}$ can be seen with an intensity of $98\pm4$ cps. However, upon increasing the power of the blocking laser we observe a pronounced decrease in the intensity of $X_{ud}$, while emission from $X_{ld}$ emerges and gradually increases in intensity.  These observations clearly demonstrate a conditional optical response, blocking of $X_{ud}$ induced by direct excitation of $X_{ld}$. To quantitatively analyze the effect of the blocking laser on the PL signal from the QDM we define the intensity of $X_{ud}$ subject to the combined pump and blocking lasers as $I_{block, pump}$: there $I_{1,1}$ corresponds to the intensity measured when blocking and pump lasers are switched on, $I_{0,1}$ is the intensity with the pump laser etc. These three intensities are plotted together with the intensity of the lower dot in Fig. 4(b) as a function of the blocking laser power. We measure $I_{0,1} = 98\pm4$ cps (red triangles) and $I_{1,1} = 19\pm1$ cps (green squares) as the intensity of the blocking laser is increased from 0 to 4 kW/cm. When only the blocking laser is applied we measure $I_{1,0}=9\pm1$ cps (blue triangles) from which we can obtain the suppression $\rho = \frac{I_{0,1}-I_{1,1}}{I_{0,1}-I_{1,0}} \sim 88\pm 2\%$. The dependence of $\rho$ on the blocking laser power is presented in Fig.4(c). The reason why the blocking laser induced PL from $X_{ud}$ results from the fact that the resonance used for efficiently exciting $X_{ld}$ is 22.3 meV higher than the ground state of $X_{ld}$. Therefore, relaxation from this excited state to the ground state of $X_{ud}$ is possible due to tunneling of both charge carriers.

In summary, we probed the spectrum of ground and excited state transitions in an individual, electrically tunable artificial molecule.  Excited state transitions were identified between hybridized electron states having bonding or antibonding character and different excited hole states. By simultaneously pumping different discrete optical transitions we demonstrated a conditional optical response with an on-off gating fidelity of $88 \pm 2 \%$. The results demonstrate an
electrically tunable, discrete coupled quantum system with a conditional optical response.

We gratefully acknowledge financial support of the DFG via SFB-631 / Nanosystems Initiative Munich and the EU via SOLID.

\bibliography{Papers}

\end{document}